\documentclass[epj]{webofc}
\usepackage[utf8]{inputenc}
\usepackage[varg]{txfonts}   % Web of Conferences font
\usepackage{booktabs}
\usepackage{xcolor}
\definecolor{darkred}{rgb}{0.4,0.0,0.0}
\definecolor{darkgreen}{rgb}{0.0,0.4,0.0}
\definecolor{darkblue}{rgb}{0.0,0.0,0.4}
\usepackage[bookmarks,linktocpage,colorlinks,
    linkcolor = darkred,
    urlcolor  = darkblue,
    citecolor = darkgreen]{hyperref}
\usepackage{subfigure}
\usepackage{multicol}
\usepackage{multirow}
\usepackage{bm}
\usepackage{caption}
\wocname{EPJ Web of Conferences}
\woctitle{Lattice2017}
\providecommand{\fm}{\mathrm{fm}}
\providecommand{\MeV}{\mathrm{MeV}}
\providecommand{\GeV}{\mathrm{GeV}}

\providecommand{\abs}[1]{\lvert#1\rvert}

\providecommand{\ket}[1]{\lvert#1\rangle}
\providecommand{\matrixe}[3]{\langle#1\lvert#2\rvert#3\rangle}
\providecommand{\expv}[1]{\langle#1\rangle}

\providecommand{\RN}[1]{%
  \textup{\uppercase\expandafter{\romannumeral#1}}%
}

\providecommand{\CL}{\nonumber\\}

\graphicspath{{./figs/}}
\allowdisplaybreaks

\begin{document}

\selectlanguage{english}

\title{%
Nucleon Axial and Electromagnetic Form Factors
}
%\author{%
%\firstname{Bilbo} \lastname{Baggins}\inst{1,3}\fnsep\thanks{Acknowledges financial support by his mentor J.R.R. Tolkien.} \and
%\firstname{Harry} \lastname{Potter}\inst{2} \and
%\firstname{Star}  \lastname{Lord}\inst{3}\fnsep\thanks{Speaker, \email{guardians@galaxy.net} (only for submitting author)}
%% etc.
%}
\author{%
  \firstname{Yong-Chull} \lastname{Jang}\inst{1}\fnsep\thanks{Speaker. \email{ypj@bnl.gov}. Present: Physics Department, Brookhaven National Laboratory, Upton, NY 11973, USA. } \and
\firstname{Tanmoy} \lastname{Bhattacharya}\inst{1} \and
\firstname{Rajan}  \lastname{Gupta}\inst{1} \and
\firstname{Huey-Wen}  \lastname{Lin}\inst{3} \and
\firstname{Boram}  \lastname{Yoon}\inst{2}
}
\institute{%
Theoretical Division T-2, 
Los Alamos National Laboratory, 
Los Alamos, NM 87545, U.S.A.
\and
Computer, Computational, and Statistical Sciences CCS-7, 
Los Alamos National Laboratory,
Los Alamos, NM 87545, U.S.A.
\and
Department of Physics and Astronomy, 
Michigan State University, MI 48824, U.S.A
}

\abstract{We present results for the isovector axial, induced
  pseudoscalar, electric, and magnetic form factors of the
  nucleon. The calculations were done using $2+1+1$-flavor HISQ
  ensembles generated by the MILC collaboration with lattice spacings
  $a \approx$ 0.12, 0.09, 0.06\,$\fm$ and pion masses $M_\pi \approx$
  310, 220, 130\,$\MeV$.  Excited-states contamination is controlled by
  using four-state fits to two-point correlators and by comparing two-
  versus three-states in three-point correlators.  The $Q^2$ behavior
  is analyzed using the model independent z-expansion and the dipole
  ansatz. Final results for the charge radii and magnetic moment are
  obtained using a simultaneous fit in $M_\pi$, lattice spacing $a$
  and finite volume.
%The same calculation with the $2+1$-flavor clover ensembles with $M_\pi \approx$ 300, 190\,$\MeV$ is compared with the results from HISQ ensemble.
}
\maketitle

\section{Introduction}\label{intro}

To extract the form factors from the three-point correlators, we consider the spectral decomposition  including
contributions from three states, the ground state $\ket{0}$ and two
excited states $\ket{1}, \ket{2}$: 
\begin{align}
  C_\Gamma^{(3\text{pt})}(t;\tau;\bm{p}^\prime,\bm{p}) &= 
  \abs{\mathcal{A}_0^\prime} \abs{\mathcal{A}_0}\matrixe{0^\prime}{\mathcal{O}_\Gamma}{0} e^{-E_0t - M_0(\tau-t)} \CL
   &\;  + \abs{\mathcal{A}_0^\prime} \abs{\mathcal{A}_1}\matrixe{0^\prime}{\mathcal{O}_\Gamma}{1} e^{-E_0t - M_1(\tau-t)}
   + \abs{\mathcal{A}_1^\prime} \abs{\mathcal{A}_0}\matrixe{1^\prime}{\mathcal{O}_\Gamma}{0} e^{-E_1t -M_0(\tau-t)} \CL
   &\;+ \abs{\mathcal{A}_1^\prime} \abs{\mathcal{A}_1}\matrixe{1^\prime}{\mathcal{O}_\Gamma}{1} e^{-E_1t - M_1(\tau-t)} \CL
   &\;  + \abs{\mathcal{A}_0^\prime} \abs{\mathcal{A}_2}\matrixe{0^\prime}{\mathcal{O}_\Gamma}{2} e^{-E_0t - M_2(\tau-t)}
   + \abs{\mathcal{A}_2^\prime} \abs{\mathcal{A}_0}\matrixe{2^\prime}{\mathcal{O}_\Gamma}{0} e^{-E_2t -M_0(\tau-t)} \CL
   &\;  + \abs{\mathcal{A}_1^\prime} \abs{\mathcal{A}_2}\matrixe{1^\prime}{\mathcal{O}_\Gamma}{2} e^{-E_1t - M_2(\tau-t)}
   + \abs{\mathcal{A}_2^\prime} \abs{\mathcal{A}_1}\matrixe{2^\prime}{\mathcal{O}_\Gamma}{1} e^{-E_2t -M_1(\tau-t)} \CL
   &\;+ \abs{\mathcal{A}_2^\prime} \abs{\mathcal{A}_2}\matrixe{2^\prime}{\mathcal{O}_\Gamma}{2} e^{-E_2t - M_2(\tau-t)} \,.
   \label{eq:3pt}
\end{align}
In our lattice calculation, $\bm{p}=\bm{0}$ and the three states have
mass $M_i$. The primed states $\ket{j^\prime}$ have momentum
$\bm{p}^\prime$ and energy $E_j$. The desired matrix element is
$\matrixe{0^\prime}{\mathcal{O}_\Gamma}{0}$, which can be decomposed into
nucleon form factors, associated with all possible Lorentz covariant
structures for a given current insertion $O_\Gamma$. To estimate
convergence of the truncated spectral decomposition, we compare
results of 2-state fits (neglecting contributions of the second
excited state) with a $3^\ast$-state fit in
which the poorly determined matrix element
$\matrixe{2^\prime}{\mathcal{O}_\Gamma}{2}$ is set to zero. Within the
single elimination jackknife process, we use results of 4-state fits
to the two-point correlator to obtain the energy $E_i$, mass $M_i$ and
amplitudes $\mathcal{A}_i^{(\prime)}$ that are inputs in the fits to the 
three-point correlators using Eq.~\eqref{eq:3pt}. 

\clearpage
\begin{table}[!tb]
  \centering
  \caption{Fit parameters. The $2^{nd}$ column gives the fit ranges used
    for nucleon two-point correlators. The $3^{rd}$  column gives the
    values of source-sink separations $\tau$ simulated and used in the
    fits, and the $4^{th}$ column gives the number of timeslices, $t_\text{skip}$,  
    adjacent to the source and the sink, skipped in the fits to
    three-point correlators to control excited-state contamination.
    The $5^{th}$ column gives the value of $\bar{t}_0$ chosen in the
    $z-$expansion fit at which $z(Q^2=\bar{t}_0)=0$. The fit ranges
    for ensembles a09m310 and a09m220 are different from those
    in~\cite{Rajan:2017lxk}, since these ensembles have been updated
    with higher statistics AMA bias corrected data, and include data
    with $\tau=16$ and momentum insertion up to
    $n^2=10$. The calculation of the $Q^2\neq 0$ data for the a09m130 ensemble has been analyized using only the low precision data.\looseness-1}
  \label{tab:fitrange}
  \renewcommand{\arraystretch}{1.0}
  \setlength{\tabcolsep}{4pt}
  \begin{tabular}{lcccc|cc|ccc}
  \hline\hline
  ensemble & $[t_\text{min},t_\text{max}]$ & $\{\tau\}$ & $t_\text{skip}$ & $\bar{t}_0$ & $L^3\times T$    & $M_\pi^{\rm val} L$ & $N_\text{conf}$  & $N_{\rm meas}^{\rm HP}$  & $N_{\rm meas}^{\rm {LP}}$  \\
  \hline
  a12m310  & $[2,15]$ & \{8,10,12\}     & 2 & 0.40 & $24^3\times 64$ & 4.55 & 1013 & 8104  &   64,832  \\
  a12m220L & $[2,15]$ & \{8,10,12,14\}  & 2 & 0.20 & $40^3\times 64$ & 5.49 & 1010 & 8080  &   68,680  \\
  a09m310  & $[2,18]$ & \{10,12,14,16\} & 3 & 0.50 & $32^3\times 96$ & 4.51 & 2264 & 9056  &  114,896  \\
  a09m220  & $[3,20]$ & \{10,12,14,16\} & 3 & 0.40 & $48^3\times 96$ & 4.79 & 964  & 3856  &  123,392  \\
  a09m130  & $[4,20]$ & \{10,12,14\}    & 3 & 0.12 & $64^3\times 96$ & 3.90 & 883  & 7064  &  56,512   \\
  a06m310  & $[7,30]$ & \{16,20,22,24\} & 7 & 0.40 & $48^3\times 144$& 4.52 & 1000 & 8000  &  64,000   \\
  a06m220  & $[7,30]$ & \{16,20,22,24\} & 7 & 0.20 & $64^3\times 144$& 4.41 & 650  & 2600  &  41,600   \\
  a06m135  & $[6,30]$ & \{16,18,20,22\} & 6 & 0.12 & $96^3\times 192$& 3.74 & 322  & 1288  &  20,608    \\
  \hline\hline
  \end{tabular}
\end{table}

\section{Axial Form Factor}

\begin{figure}[thb] % no figure before 1st section
  \centering
  \subfigure[2-state]{%
    \includegraphics[width=0.46\textwidth,clip]{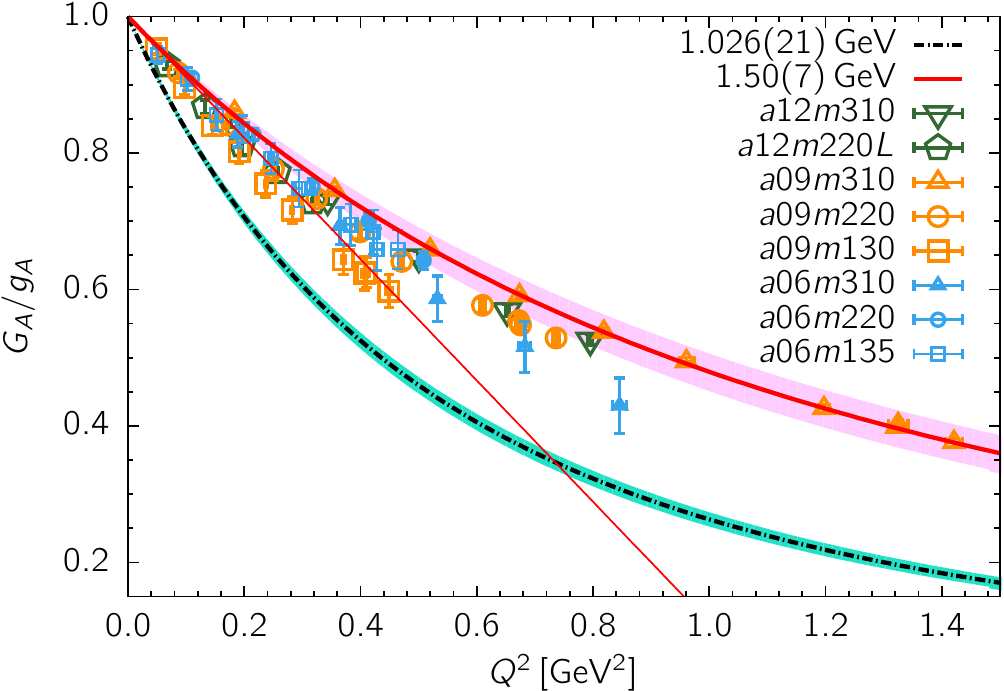}
  }\hfill
  \subfigure[$3^\ast$-state]{%
    \includegraphics[width=0.46\textwidth,clip]{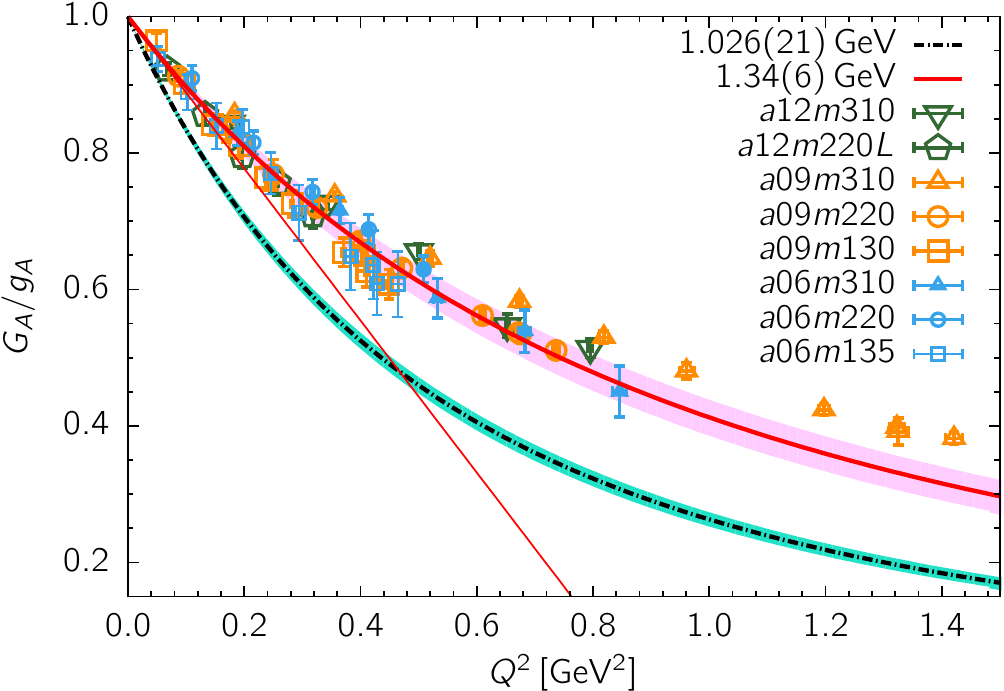}
  }
  \caption{Axial form factor data from the 2- and $3^\ast$-state fits
    to the three-point functions.  The thick red line within the pink
    band shows the dipole result given in Table~\ref{tab:final}, and
    the thin straight red line is the slope, $-r_A^2/6$, at zero
    momentum transfer. Both the $z-$expansion and the dipole estimates
  differ from the dipole fit with the phenominological estimate,
    $\mathcal{M}_A=1.026(21)\,\GeV$ given in
    Ref.~\cite{Bernard:2001rs}, that represents the world average from the 
    neutrino scattering data.  Note also the change in the slope,
    $-r_A^2/6$, between the 2- and $3^\ast$-state fits. This trend is the same for the $z$-expansion estimates.}
  \label{fig:GA}
\vskip-30pt
\end{figure}

Nucleon matrix elements with the insertion of the isovector axial vector current
can be decomposed into the axial form factor $G_A$ and the induced
pseudoscalar form factor $\tilde{G}_P$:
\begin{align}
\label{eq:AFFdef}
\left\langle N(\bm{p}_f) | A_\mu (\bm{q}) | N(\bm{p}_i)\right\rangle &=\; 
{\overline u}_N(\bm{p}_f)\left( G_A(Q^2) \gamma_\mu
+ q_\mu \frac{\tilde{G}_P(Q^2)}{2 M_N}\right) \gamma_5 u_N(\bm{p}_i),
\end{align}
where $Q^2\equiv \bm{p}^2-(E-m)^2 = -q^2$ and $q=p_f-p_i$. Note that
$\bm{p}_i=0$ in our lattice calculation. Results for the axial form
factor $G_A(Q^2)$, normalized by the corresponding $g_A \equiv G_A(0)$
for each of the 8 ensembles, are shown in Fig.~\ref{fig:GA}. A notable
change on going from 2-state fits presented in
Ref.~\cite{Rajan:2017lxk} to $3^\ast$-state fits is the much better
agreement in the data from the two physical mass ensembles and in the
final estimates given in Table~\ref{tab:final}.  For each ensemble,
the axial charge radius $\expv{r_A^2}$ is obtained from the analytic
derivative of the dipole and the $z$-expansion fits evaluated at
$Q^2=0$ as explained in Ref.~\cite{Rajan:2017lxk}.

The chiral, continuum, and finite volume (FV) extrapolation to $M_\pi \to 135\;\MeV$,
$a\to0$ and $M_\pi L \to \infty$ is performed using only the leading 
order correction terms:
\begin{equation}
  \expv{r_A^2}(a,M_\pi,L) = c_1^A + c_2^A a + c_3^A M_\pi^2 + c_4^A M_\pi^2 \exp(-M_\pi L)
  \label{eq:rAsq-extrap}
\end{equation}
In all the results presented in this talk, the FV term is small and
$c^{A,S,T}_4$ are not well determined. Nevertheless, results with and
without the FV term are consistent as shown in Tables~\ref{tab:rA},
~\ref{tab:rE} and~\ref{tab:rM} where we give results with and without
the FV term, compare the 2- and $3^\ast$-state fits used to control
excited-state contamination, and the $z-$expansion versus the dipole
fits for the $Q^2$ behavior. A detailed description of our analysis
methodology is presented in Ref.~\cite{Rajan:2017lxk} for the axial
form factor.

For our final estimates summarized in Table~\ref{tab:final}, we
separately quote the weighted average of the two $z$-expansion fits
and the dipole results given in Table~\ref{tab:rA} including the
finite volume term.  We also quote $\mathcal{M}_A^2 \equiv 12/r_A^2$ for both
the dipole and the $z-$expansion data.  

These results are consistent with our previously reported values in
Ref.~\cite{Rajan:2017lxk}. Our new central values from the $3^\ast$-state fit 
agree with the MiniBooNE results
$\mathcal{M}_A=1.35(17)\;\GeV$~\cite{AguilarArevalo:2010zc}, but
differ by about $1\,\sigma$ from the 2-state fit results and by about
$2.5\,\sigma$ from the phenomenological estimate
$r_A=0.666(17)\,\fm$~\cite{Bernard:1998gv} obtained using the neutrino
scattering data. A recent reanalysis of the deuterium data based on
the $z$-expansion assesses an order of magnitude larger error,
$r_A=0.68(16)\,\fm$~\cite{Meyer:2016oeg}, in which case the
disagreement with our $3^\ast$-state result reduces to about
$1\sigma$.

%% %
%% % a + mpisq + fv1
%% % take positive 100% correlation for average
%% \begin{align}
%%   r_A &= 0.44(5)\;\fm\,, \quad \mathcal{M}_A = 1.56(18)\;\GeV \quad : z-\text{expansion} + 2\text{-state}\,,\CL
%%   r_A &= 0.50(6)\;\fm\,, \quad \mathcal{M}_A = 1.36(17)\;\GeV \quad : z-\text{expansion} + 3^\ast\text{-state}\,,
%%   \label{eq:rA-z}
%% \end{align}
%% %
%% \begin{align}
%%   r_A &= 0.46(2)\;\fm\,, \quad \mathcal{M}_A = 1.50(07)\;\GeV \quad : \text{dipole} + 2\text{-state}\,,\CL
%%   r_A &= 0.51(2)\;\fm\,, \quad \mathcal{M}_A = 1.34(06)\;\GeV \quad : \text{dipole} + 3^\ast\text{-state}\,.
%%   \label{eq:rA-d}
%% \end{align}
%% %
%% 

\begin{table}[!tbh]
  \centering
  \caption{Estimates for the mean square axial charge radius
    $\expv{r_A^2}$. The first column lists the terms kept in the
    continuum chiral extrapolation fit using
    Eq.~\protect\eqref{eq:rAsq-extrap}.  Data from each of the eight
    ensembles described in Table~\ref{tab:fitrange} are analyzed using 
    both the $2$-state and $3^\ast$-state truncation of the spectral decomposition of
    the three-point correlator given in Eq.~\protect\eqref{eq:3pt}, 
    followed by dipole and $z$-expansion (including sumrule
    constraints) fits described in Ref.~\cite{Rajan:2017lxk}.}
  \label{tab:rA}
  \renewcommand{\arraystretch}{1.1}
  \begin{tabular}{l|ccc|ccc}
  \hline\hline
  & \multicolumn{3}{c|}{$2$-state} & \multicolumn{3}{c}{$3^\ast$-state} \\
  \cline{2-7}
  & dipole & $z^{2+4}$ & $z^{3+4}$ & dipole & $z^{2+4}$ & $z^{3+4}$ \\
  \hline
  $a$, $M_\pi^2$, FV  & 0.208(19) & 0.180(37) & 0.223(60) & 0.260(25) & 0.245(52) & 0.272(89) \\
  %chsq/dof $a$, $M_\pi^2$, FV  & 1.51 & 1.14 & 0.78 & 1.39 & 0.24 & 0.89 \\
  $a$, $M_\pi^2$      & 0.214(15) & 0.166(29) & 0.172(48) & 0.248(20) & 0.219(46) & 0.219(79) \\
  %chisq/dof $a$, $M_\pi^2$     & 1.26 & 0.99 & 1.02 & 1.24 & 0.42 & 1.04 \\
  \hline\hline
  \end{tabular}
\vskip-20pt
\end{table}

\section{Pseudoscalar Form Factor}

\begin{figure}[thb] % no figure before 1st section
  \centering
  \subfigure[2-state]{%
    \includegraphics[width=0.46\textwidth,clip]{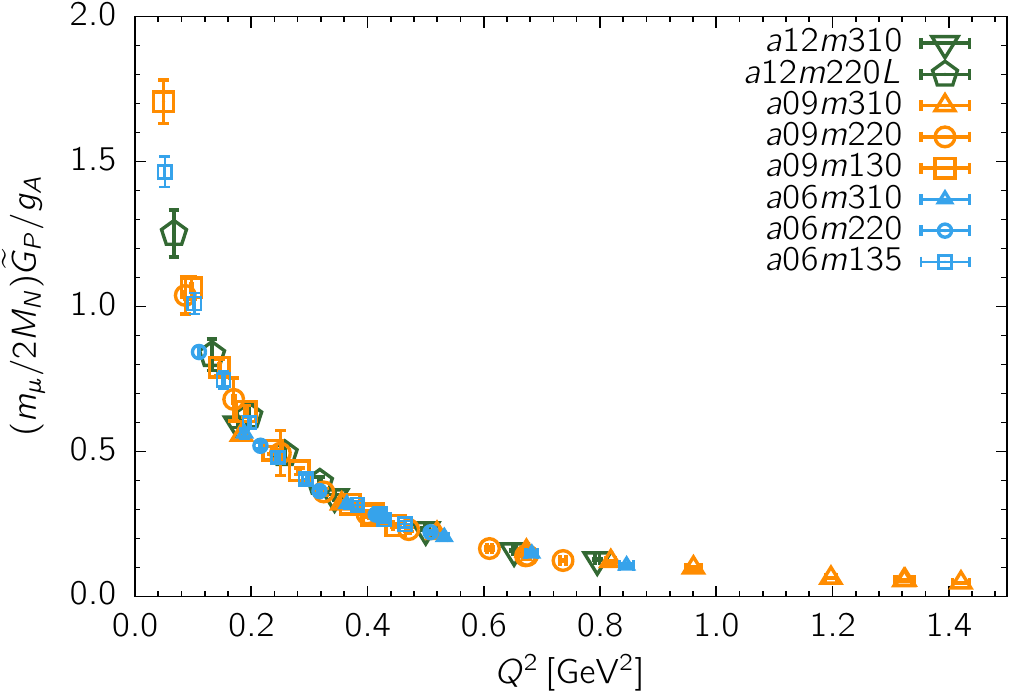}
  }\hfill
  \subfigure[$3^\ast$-state]{%
    \includegraphics[width=0.46\textwidth,clip]{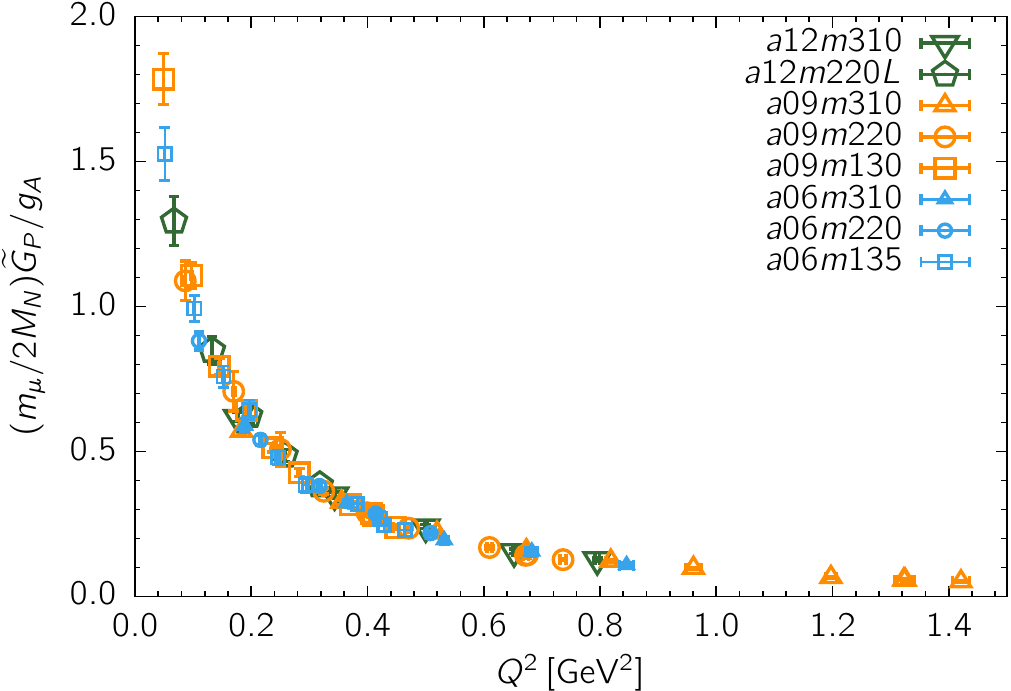}
  }
  \caption{Data for the normalized pseudoscalar form factor $(m_\mu/2M_N) \tilde{G}_P/g_A$ for the 8 ensembles.}
  \label{fig:GP}
\vskip-20pt
\end{figure}

Data for the normalized induced pseudoscalar form factor,
$(m_\mu/2M_N) \tilde{G}_P/g_A$ with $m_\mu$ the muon mass, 
% defined in Eq.~\eqref{eq:AFFdef}, 
are summarized in Fig.~\ref{fig:GP}. They show essentially no dependence on 
$M_\pi$ or $a$ or $M_\pi L$.  In
Ref.~\cite{Rajan:2017lxk}, we highlighted a problem in the extraction
of $\tilde{G}_P$: the three form factors $G_A$, $\tilde{G}_P$, and the
pseudoscalar form factor $G_P$ do not satisfy the axial Ward
identity. As a result, the pion-pole dominance ansatz used to
extrapolate the lattice data for $\tilde{G}_P$ at fixed $Q^2$ in
$M_\pi^2$, to obtain say $g_P^\ast \equiv m_\mu/2M_N \times
\tilde{G}_P(Q^2=0.88m_\mu^2)$, was shown to also fail.  In fact, our
results for $g_P^\ast$ from the two physical pion mass ensembles are 
about half the muon capture experiment result~\cite{Rajan:2017lxk}.  A
similar underestimate also occurs for the pion-nucleon coupling
$g_{\pi {\rm NN}}$.  In Ref.~\cite{Rajan:2017lxk}, we further show
that $\mathcal{O}(a)$ improvement of the axial current operator does
not significantly reduce the problem. Updated data presented here in
Fig.~\ref{fig:GP}, show only a small increase in the values of the
form factor at low $Q^2$ on going from the 2-state to the
$3^\ast$-state analysis. Thus, the violation of PCAC in
the extraction of $\tilde{G}_P$ remains an unresolved
problem. \looseness-1

\section{The Electric Form Factor}

\begin{figure}[thb] % no figure before 1st section
  \centering
  \subfigure[2-state]{%
    \includegraphics[width=0.46\textwidth,clip]{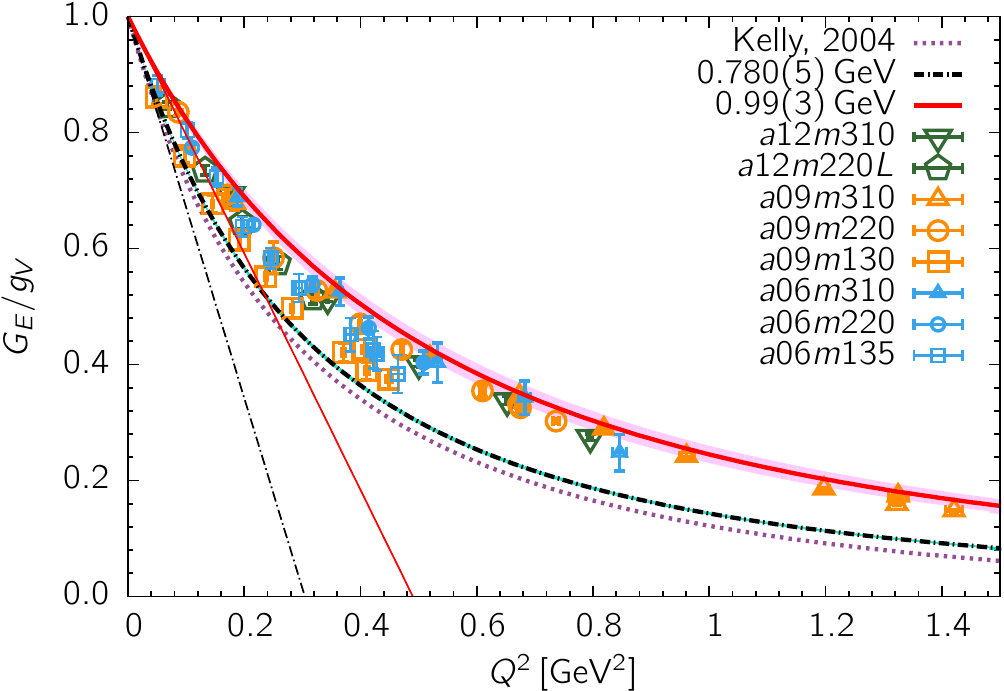}
  }\hfill
  \subfigure[$3^\ast$-state]{%
    \includegraphics[width=0.46\textwidth,clip]{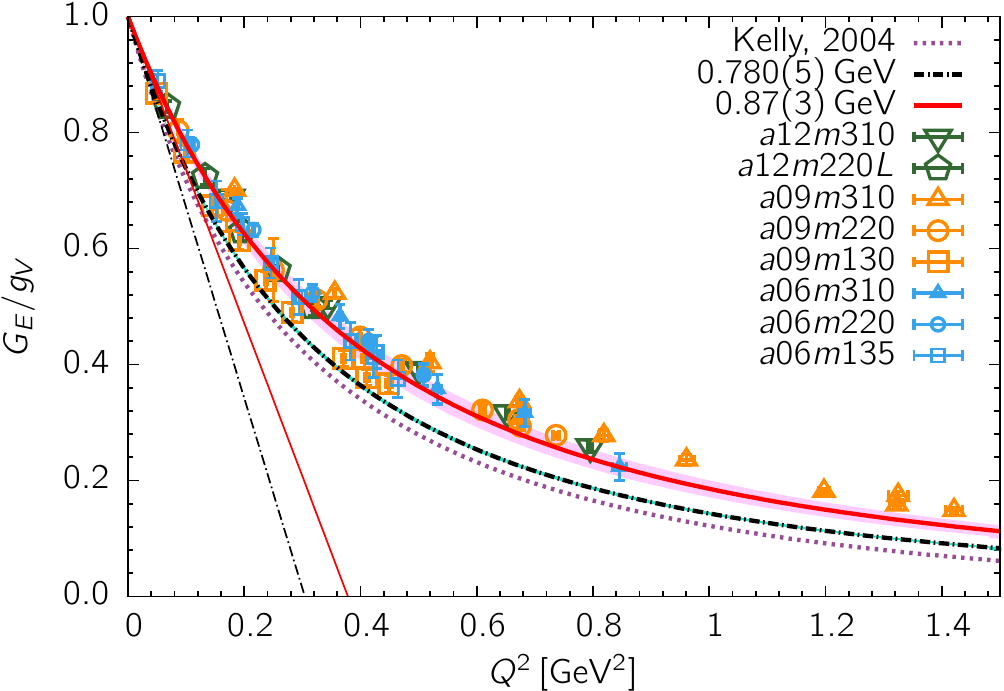}
  }
  \caption{The 8 ensemble data for the normalized electric form factor
    ${G}_E/g_V$.  The overlaid red band shows our dipole result given
    in Table~\protect\ref{tab:final}. The black dashed line shows the
    phenomenological value $\mathcal{M}_E=0.780(5)$ in both
    panels. The corresponding straight lines give their slopes,
    $-r_E^2/6$, at $Q^2 = 0$. Experimental data paramterized by the
    Kelly curve is shown by the purple dotted line.}
  \label{fig:GE}
\vskip-20pt
\end{figure}

Nucleon matrix elements with vector current insertion can be decomposed
into the Dirac and Pauli form factors $F_1$ and $F_2$ as:
\begin{align}
\label{eq:VFFdef}
\left\langle N(\bm{p}_f) | V_\mu (\bm{q}) | N(\bm{p}_i)\right\rangle &=\; 
{\overline u}_N(\bm{p}_f)\left( F_1(Q^2) \gamma_\mu
+ \sigma_{\mu\nu} \frac{F_2(Q^2)}{2 M_N}\right) \gamma_5 u_N(\bm{p}_i).
\end{align}
Here, we present results for the related Sachs, the electric and the magnetic, form factors $G_E$ and $G_M$:
\begin{align}
  G_E(Q^2) &= F_1(Q^2) - \frac{Q^2}{4M_N^2} F_2(Q^2) \,,
  \label{eq:GE} \\
  G_M(Q^2) &= F_1(Q^2) + F_2(Q^2) \,.
  \label{eq:GM}
\end{align}
The data for $G_E(Q^2)$ is summarized in Fig.~\ref{fig:GE}, and we find that 
the $3^\ast$-state fits are closer to the phenomenological curve compared to 
the $2$-state fits.  
The charge radii $\expv{r_E^2}$ and
$\expv{r_M^2}$ on each ensemble are then extracted following the same
procedure as for $\expv{r_A^2}$. From these, the continuum chiral
extrapolation for the electric charge radius is performed using the
following ansatz: 
\begin{equation}
%%  \expv{r_E^2}(a,M_\pi,L) = c_1^E + c_2^E a + c_3^E \ln(M_\pi^2/\lambda^2) + c_4^E \FV{1} \,,
  \expv{r_E^2}(a,M_\pi,L) = c_1^E + c_2^E a + c_3^E \ln(M_\pi^2/\lambda^2) + c_4^E \ln(M_\pi^2/\lambda^2) \exp(-M_\pi L) \,,
  \label{eq:rEsq-extrap}
\end{equation}
where the mass scale $\lambda$ is chosen to be $M_\rho=775\,\MeV$ and
the form of the chiral and FV corrections are taken from Refs.~\cite{Bernard:1998gv,Gockeler:2003ay}:
Using Eq.~\eqref{eq:rEsq-extrap}, the results for the different fit ansatz are summarized in Table~\ref{tab:rE}. 
For the final estimates given in Table~\ref{tab:final}, we take the weighted average 
of the two $z$-expansion fits given in Table~\ref{tab:rE}. 
The $z$-expansion and the dipole fit results with the $3^\ast$-state
analysis overlap. All four estimates are smaller than the
CODATA-2014 world average, $r_E=0.875(6)\,\fm$~\cite{Mohr:2015ccw},
from the electron experiments and the more accurate value derived from
the Lamb shift in muonic hydrogen,
$r_E=0.8409(4)\,\fm$~\cite{Antognini:2015moa}; the $z$-expansion result with $3^\ast$-state analysis is consistent with the experiments because of the error estimate is larger.

%% % a + log(mpisq) + fv1
%% % take positive 100% correlation for average
%% \begin{align}
%%   r_E &= 0.78(5)\;\fm\,, \quad \mathcal{M}_E = 0.87(5)\;\GeV \quad : z-\text{expansion} + 2\text{-state}\,,\CL
%%   r_E &= 0.79(5)\;\fm\,, \quad \mathcal{M}_E = 0.87(6)\;\GeV \quad : z-\text{expansion} + 3^\ast\text{-state}\,,
%%   \label{eq:rE-z}
%% \end{align}
%% %
%% and from the dipole fits we get
%% %
%% \begin{align}
%%   r_E &= 0.73(1)\;\fm\,, \quad \mathcal{M}_E = 0.94(2)\;\GeV \quad : \text{dipole} + 2\text{-state}\,,\CL
%%   r_E &= 0.77(2)\;\fm\,, \quad \mathcal{M}_E = 0.89(2)\;\GeV \quad : \text{dipole} + 3^\ast\text{-state}\,.
%%   \label{eq:rE-d}
%% \end{align}
%% % a + mpisq + fv1
%% % take positive 100% correlation for average
%% %\begin{align}
%% %  r_E &= 0.72(4)\;\fm\,, \quad \mathcal{M}_E = 0.95(6)\;\GeV \quad : z-\text{expansion} + 2\text{-state}\,,\\
%% %  r_E &= 0.77(6)\;\fm\,, \quad \mathcal{M}_E = 0.89(6)\;\GeV \quad : z-\text{expansion} + 3^\ast\text{-state}\,,\\
%% %  r_E &= 0.62(2)\;\fm\,, \quad \mathcal{M}_E = 1.01(3)\;\GeV \quad : \text{dipole} + 2\text{-state}\,,\\
%% %  r_E &= 0.73(2)\;\fm\,, \quad \mathcal{M}_E = 0.93(3)\;\GeV \quad : \text{dipole} + 3^\ast\text{-state}\,.
%% %  \label{eq:rE}
%% %\end{align}
%% %

\begin{table}[!tb]
  \centering
  \caption{Mean square electric charge radius $\expv{r_E^2}$. The
    first column shows the terms included in the chiral continuum
    extrapolation defined in Eq.~\eqref{eq:rAsq-extrap}. 
    %\ypj{The $z$-expansion fit with $3^\ast$-state fit gives a reasonable $\chi^2$/d.o.f $\sim$ 0.5--1.0, whereas other fits result about 2--5.}{} 
    The rest is the same as in Table~\ref{tab:rA}.}
  \label{tab:rE}
  \renewcommand{\arraystretch}{1.1}
  \begin{tabular}{l|ccc|ccc}
  \hline\hline
  & \multicolumn{3}{c|}{$2$-state} & \multicolumn{3}{c}{$3^\ast$-state} \\
  \cline{2-7}
  & dipole & $z^{2+4}$ & $z^{3+4}$ & dipole & $z^{2+4}$ & $z^{3+4}$ \\
  \hline
  %$a$, $\ln M_\pi^2$, FV1    & 0.529(21) & 0.550(62) & 0.807(109) & 0.586(31) & 0.560(71) & 0.742(106) \\
  $a$, $\ln M_\pi^2$, FV    & 0.473(32) & 0.475(83) & 0.529(160) & 0.619(49) & 0.638(124) & 0.801(174) \\
  %chisq/dof $a$, $\ln M_\pi^2$, FV    & 4.62 & 2.77 & 2.28 & 2.03 & 0.99 & 0.58 \\
  $a$, $\ln M_\pi^2$        & 0.531(21) & 0.528(54) & 0.730(097) & 0.580(30) & 0.561(071) & 0.738(105) \\
  %chisq/dof $a$, $\ln M_\pi^2$        & 4.89 & 2.36 & 2.33 & 1.83 & 0.91 & 0.51 \\
%%  $a$, $M_\pi^2$, $\FV{1}$   & 0.455(25) & 0.472(53) & 0.661(101) & 0.537(36) & 0.552(73) & 0.680(111) \\
%%  $a$, $M_\pi^2$ & 0.506(20) & 0.497(46) & 0.722(082)  & 0.552(29) & 0.528(60) & 0.680(093) \\
  \hline\hline
  \end{tabular}
\end{table}

\section{The Magnetic Form Factor}

\begin{figure}[thb] % no figure before 1st section
  \centering
  \subfigure[2-state]{%
    \includegraphics[width=0.46\textwidth,clip]{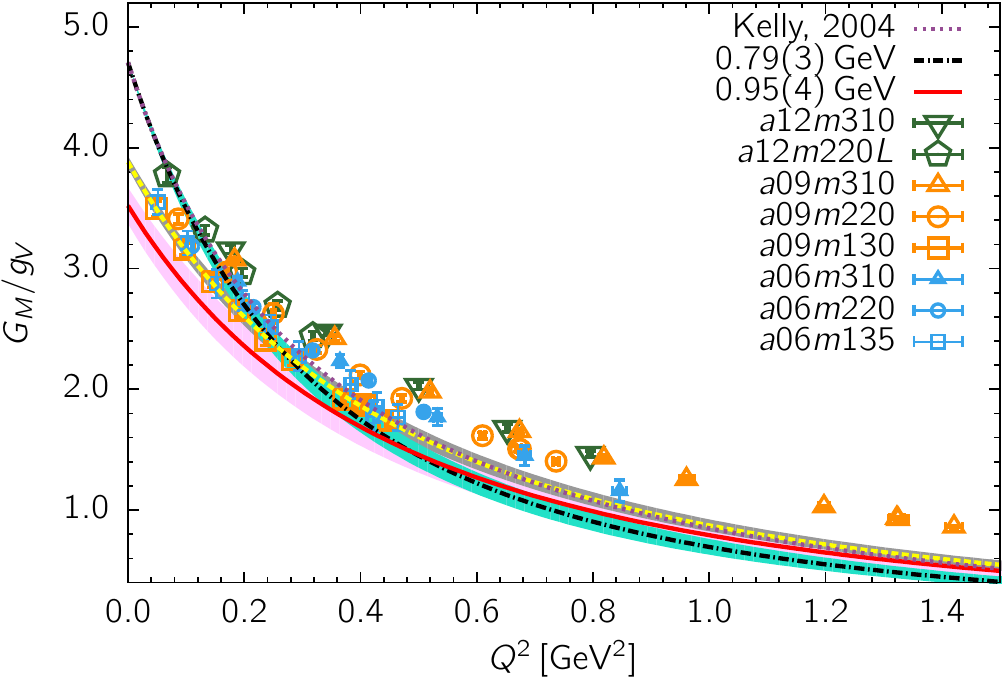}
  }\hfill
  \subfigure[$3^\ast$-state]{%
    \includegraphics[width=0.46\textwidth,clip]{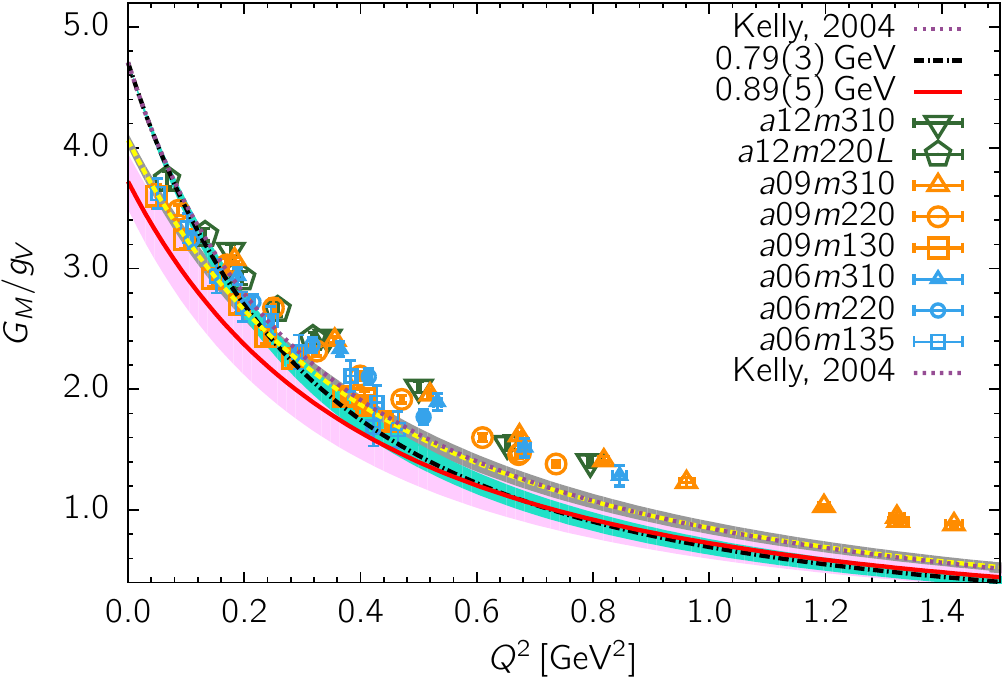}
  }
  \caption{Data for the normalized magnetic form factor ${G}_M/g_V$
    for the 8 ensembles. The yellow dashed line is the dipole fit data
    extrapolated using only a fit in $m_\pi^2$. The rest is the same as in
    Fig.~\protect\ref{fig:GE}.}
  \label{fig:GM}
\end{figure}

The $z-$expansion fits to $G_M(Q^2)$ are much less stable
since the point $F_2(Q^2=0)$ cannot be extracted from
Eq.~\eqref{eq:VFFdef}; it is obtained from the fit in $Q^2$. As a
result, the $z-$expansion estimates in Table~\ref{tab:rM} are only
with terms up to $z^3$. Results of fits with sumrules are even less
stable and not presented here. Using the data from the 8 ensembles, we
perform the continuum-chiral extrapolations for the magnetic charge
radius $r_M$ and the magnetic moment $\mu$ using the ansatz:
\begin{equation}
%%  \expv{r_M^2}(a,M_\pi,L) &=& c_1^M + c_2^M a + c_3^M/M_\pi + c_4^M FV{1} \,, \\
  \expv{r_M^2}(a,M_\pi,L) = c_1^M + c_2^M a + c_3^M/M_\pi + c_4^M/M_\pi \exp(-M_\pi L)\,, 
  \label{eq:rMsq-extrap}
\end{equation}
%
%and
%
\begin{equation}
%%  \mu(a,M_\pi,L) &=& c_1^\mu + c_2^\mu a + c_3^\mu M_\pi + c_4^\mu \FV{3}\,,\quad \FV{3} \equiv M_\pi\left(1-\frac{2}{M_\pi L}\right) \exp(-M_\pi L) \,. 
  \mu(a,M_\pi,L) = c_1^\mu + c_2^\mu a + c_3^\mu M_\pi + c_4^\mu M_\pi\left(1-\frac{2}{M_\pi L}\right) \exp(-M_\pi L) \,. 
  \label{eq:mu-extrap}
\end{equation}
The form of the chiral and FV correction terms in $\expv{r_M^2}$ are
taken from Ref.~\cite{Bernard:1998gv}. The FV term in $\mu$ is taken
from Ref.~\cite{Beane:2004tw}. The NLO chiral correction in $\mu$ has
a known coefficient, $ (g_A^2 M_N)/(4 \pi F_\pi^2) M_\pi( 1 +
(3M_\pi)/(M_N) \ln(M_\pi^2/\lambda^2) )$~\cite{Meissner:1997hn},
however, there is an additional chiral log at the same order, i.e.,
proportional to $M_\pi^2$, that involves unknown LEC. To include both chiral logs, an
additional parameter is needed. Since we have data
over a limited range of $M_\pi^2$ and with essentially three values of
$M_\pi^2$, we neglect the chiral log corrections. 
For the same reason, we also leave $c_3^\mu$ a free parameter 
rather than take the form predicted by $\chi$PT.  The results of
the fits, with and without the respective FV correction term, are
summarized in Table~\ref{tab:rM}.

\begin{table}[!tbh]
  \centering
  \caption{Results of fits for the mean square magnetic charge radius
    $\expv{r_M^2}$ using Eq.~\protect\ref{eq:rMsq-extrap} (upper
    half), and for $\mu$ using Eq.~\protect\ref{eq:mu-extrap} (lower
    half).  The second column shows the terms included in the chiral
    continuum extrapolation. The rest is the same as in
    Table~\ref{tab:rA}.}
  \label{tab:rM}
  \renewcommand{\arraystretch}{1.1}
  \begin{tabular}{c|l|ccc|ccc}
  \hline\hline
  &  & \multicolumn{3}{c|}{$2$-state} & \multicolumn{3}{c}{$3^\ast$-state} \\
  \cline{3-8}
  & & dipole & $z^{2}$ & $z^{3}$ & dipole & $z^{2}$ & $z^{3}$ \\
  \hline
%$\expv{r_M^2}$ & $a$, $M_\pi^{-1}$, FV1  & 0.463(27) & 0.575(70) & 0.325(291) & 0.490(39) & 0.566(101) & 0.512(448) \\
  \multirow{2}{*}{$\expv{r_M^2}$} & $a$, $M_\pi^{-1}$, FV  & 0.517(46) & 0.716(96) & 0.994(405) & 0.587(68) & 0.666(136) & 0.878(649) \\
               &  $a$, $M_\pi^{-1}$     & 0.468(26) & 0.619(60) & 0.483(278) & 0.477(39) & 0.591(093) & 0.580(439) \\
  %chisq/dof \multirow{2}{*}{$\expv{r_M^2}$} & $a$, $M_\pi^{-1}$, FV  & 0.67 & 0.18 & 0.55 & 0.90 & 0.44 & 0.74 \\
  %             &  $a$, $M_\pi^{-1}$     & 0.90 & 0.49 & 1.04 & 1.49 & 0.46 & 0.67 \\
%%  $a$, $M_\pi^2$, $\FV{1}$    & 0.411(30) & 0.519(55) & 0.684(280) & 0.433(46) & 0.460(087) & 0.534(473) \\
%%  $a$, $M_\pi^2$         & 0.420(25) & 0.513(48) & 0.497(241) & 0.439(37) & 0.470(078) & 0.532(390) \\
  \hline
  \multirow{2}{*}{$\mu$}          &  $a$, $M_\pi$, FV      & 3.52(15) & 3.39(19) & 3.72(42) & 3.72(23) & 3.39(30) & 3.92(70) \\
               &  $a$, $M_\pi$          & 3.48(10) & 3.41(13) & 3.34(30) & 3.64(14) & 3.49(20) & 3.63(46) \\
  %chisq/dof \multirow{2}{*}{$\mu$}          &  $a$, $M_\pi$, FV      & 1.02 & 0.55 & 0.80 & 0.77 & 0.41 & 0.59 \\
  %             &  $a$, $M_\pi$          & 0.84 & 0.44 & 0.99 & 0.65 & 0.37 & 0.53 \\
  \hline\hline
  \end{tabular}
\end{table}

%% \begin{table}[!bth]
%%   \centering
%%   \caption{Magnetic moment $\mu$. The first column shows the terms included in the chiral continuum extrapolation. See Eq.~\eqref{eq:mu-extrap} and Eq.~\eqref{eq:rAsq-extrap}. Rest is the same as in Table~\ref{tab:rA}.}
%%   \label{tab:mu}
%%   \renewcommand{\arraystretch}{1.1}
%%   \begin{tabular}{l|ccc|ccc}
%%   \hline\hline
%%   & \multicolumn{3}{c|}{$2$-state} & \multicolumn{3}{c}{$3^\ast$-state} \\
%%   \cline{2-7}
%%   & dipole & $z^{2}$ & $z^{3}$ & dipole & $z^{2}$ & $z^{3}$ \\
%%   \hline
%%   $a$, $M_\pi$, $\FV{3}$      & 3.52(15) & 3.39(19) & 3.72(42) & 3.72(23) & 3.39(30) & 3.92(70) \\
%%   $a$, $M_\pi$                & 3.48(10) & 3.41(13) & 3.34(30) & 3.64(14) & 3.49(20) & 3.63(46) \\
%%   $a$, $M_\pi^2$, $\FV{1}$    & 3.55(13) & 3.44(17) & 3.72(38) & 3.72(20) & 3.43(27) & 3.89(64) \\
%%   $a$, $M_\pi^2$              & 3.49(10) & 3.43(13) & 3.36(30) & 3.65(14) & 3.49(19) & 3.64(46) \\
%%   \hline\hline
%%   \end{tabular}
%% \end{table}
%  $a$, $M_\pi$, $\FV{3}$      & 3.52(15) & 3.39(19) & 3.72(42) & 3.48(10) & 3.41(13) & 3.34(30) \\
%  $a$, $M_\pi$                & 3.72(23) & 3.39(30) & 3.92(70) & 3.64(14) & 3.49(20) & 3.63(46) \\
%  $a$, $M_\pi^2$, $\FV{1}$    & 3.55(13) & 3.44(17) & 3.72(38) & 3.72(20) & 3.43(27) & 3.89(64) \\
%  $a$, $M_\pi^2$              & 3.49(10) & 3.43(13) & 3.36(30) & 3.65(14) & 3.49(19) & 3.64(46) \\

Our final results collected in Table~\ref{tab:final}, are obtained by
fitting $\expv{r_M^2}$ and $\mu$ using Eq.~\eqref{eq:rMsq-extrap} and
Eq.~\eqref{eq:mu-extrap}, respectively, and keeping all four terms.
For the $z-$expansion, we take a weighted average of the $z^2$ and 
$z^3$ truncation results.
Estimates from the $2-$ and $3^\ast$-state fits are consistent for
both the dipole and the $z$-expansion ansatz, but with larger errors than in 
$\expv{r_E^2}$.  The $z$-expansion gives larger central values and errors
compared to the dipole fits. The dipole estimates are smaller than the experimental value
$r_M=0.86(3)\,\fm$~\cite{Mohr:2015ccw} obtained from electron
scattering experiments but the $z$-expansion estimates are consistent with the experimental value. Nevertheless, all four estimates of $\mu$ are $3/4$ of the precisely known value $\mu=1+\kappa_p-\kappa_n=4.7058$ with 
the anomalous magnetic moments of proton $\kappa_p=1.7928$ and of the neutron
$\kappa_n=-1.9130$~\cite{Patrignani:2016xqp}.
%{Nevertheless, all four estimates in
%Table~\ref{tab:final} are smaller than the experimental value
%$r_M=0.86(3)\,\fm$~\cite{Mohr:2015ccw} obtained from electron
%scattering experiments and $\mu$ is $3/4$ of the precisely known value
%$4.7058$.}

%% %
%% % a + mpiinv + fv1
%% % take positive 100% correlation for average
%% \begin{align}
%%   r_M &= 0.75(6)\,\fm\,, \quad \mathcal{M}_M = 0.91(07)\,\GeV \quad : z-\text{expansion} + 2\text{-state}\,,\CL
%%   r_M &= 0.75(8)\,\fm\,, \quad \mathcal{M}_M = 0.91(10)\,\GeV \quad : z-\text{expansion} + 3^\ast\text{-state}\,,
%%   \label{eq:rM-z}
%% \end{align}
%% %
%% and the dipole fit gives
%% %
%% \begin{align}
%%   r_M &= 0.68(2)\,\fm\,, \quad \mathcal{M}_M = 1.00(03)\,\GeV \quad : \text{dipole} + 2\text{-state}\,,\CL
%%   r_M &= 0.70(3)\,\fm\,, \quad \mathcal{M}_M = 0.98(04)\,\GeV \quad : \text{dipole} + 3^\ast\text{-state}\,.
%%   \label{eq:rM-d}
%% \end{align}
%% % a + mpisq + fv1
%% % take positive 100% correlation for average
%% %\begin{align}
%% %  r_M &= 0.73(4)\;\fm\,, \quad \mathcal{M}_M = 0.94(06)\;\GeV \quad : z-\text{expansion} + 2\text{-state}\,,\\
%% %  r_M &= 0.68(7)\;\fm\,, \quad \mathcal{M}_M = 1.01(11)\;\GeV \quad : z-\text{expansion} + 3^\ast\text{-state}\,,\\
%% %  r_M &= 0.64(2)\;\fm\,, \quad \mathcal{M}_M = 1.07(04)\;\GeV \quad : \text{dipole} + 2\text{-state}\,,\\
%% %  r_M &= 0.66(4)\;\fm\,, \quad \mathcal{M}_M = 1.04(06)\;\GeV \quad : \text{dipole} + 3^\ast\text{-state}\,.
%% %  \label{eq:rM}
%% %\end{align}
%% %

\begin{figure}[thb] 
  \centering
  %\subfigure[]{%
    \includegraphics[width=0.46\textwidth,clip]{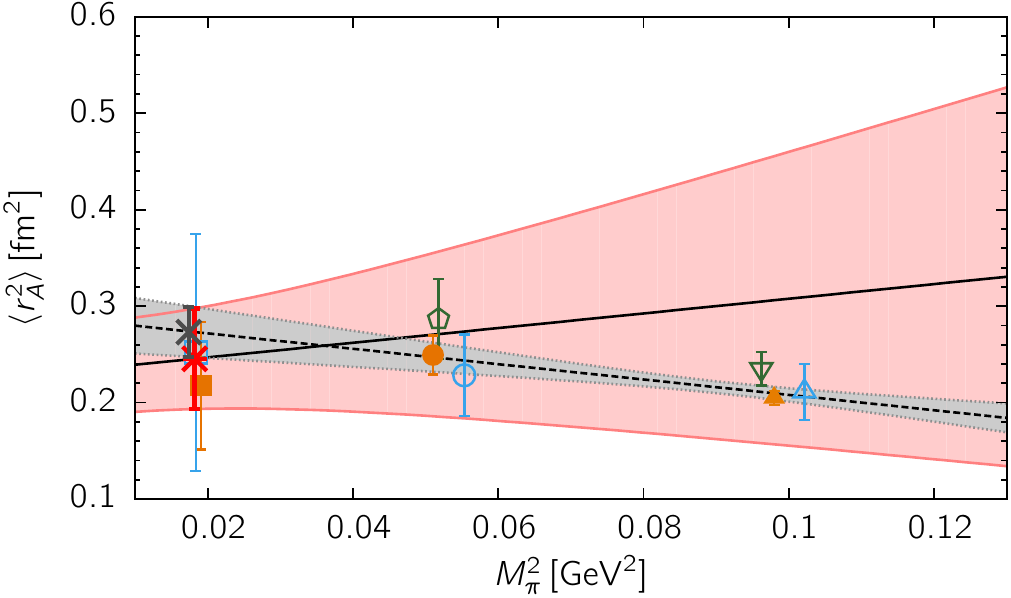}
  %}
  \hfill
  %\subfigure[]{%
    \includegraphics[width=0.46\textwidth,clip]{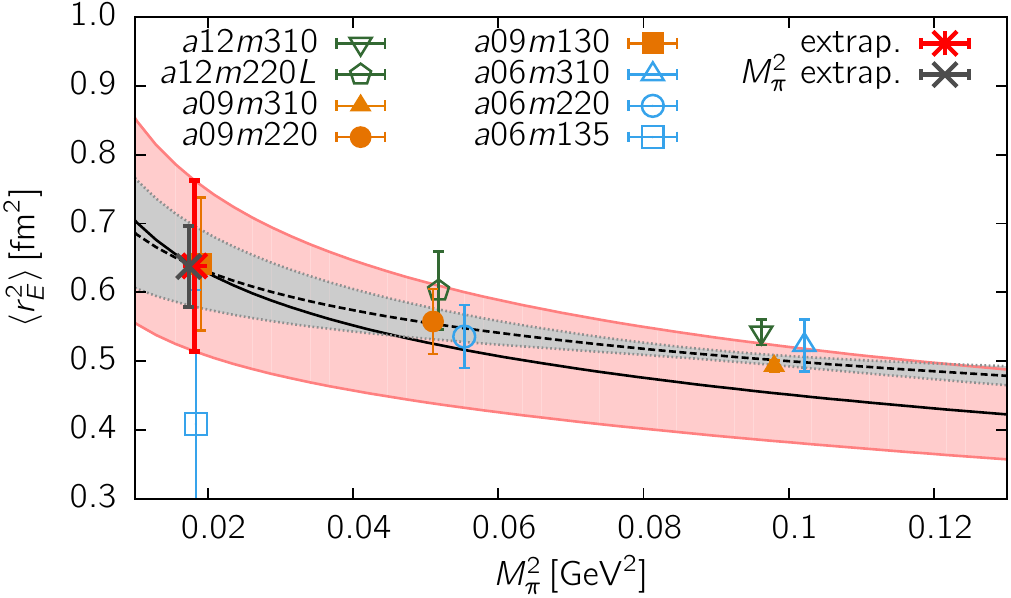}
  %}
    \\\vspace{1em}
  %\subfigure[]{%
    \includegraphics[width=0.46\textwidth,clip]{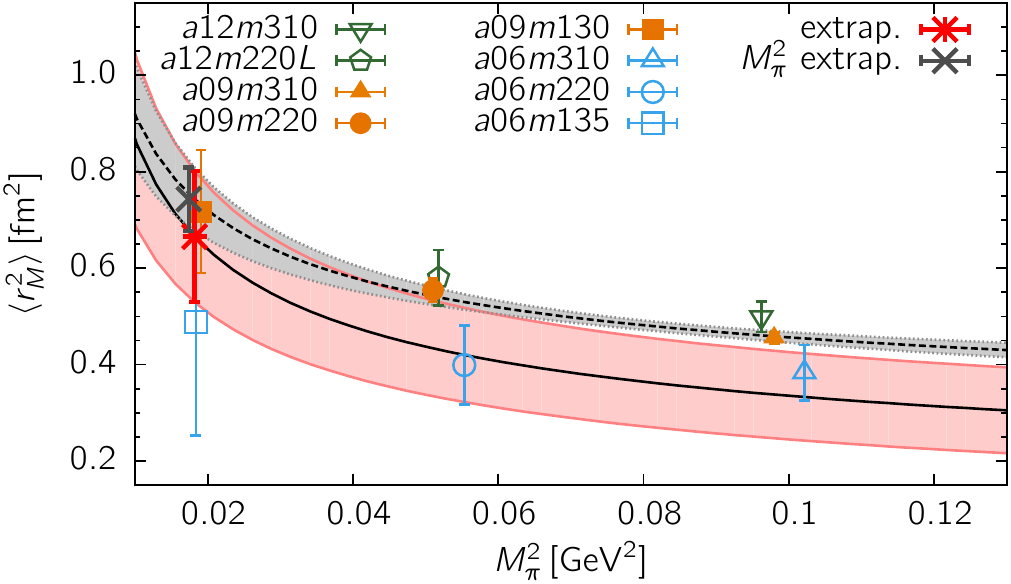}
  %}
  \hfill
  %\subfigure[]{%
    \includegraphics[width=0.46\textwidth,clip]{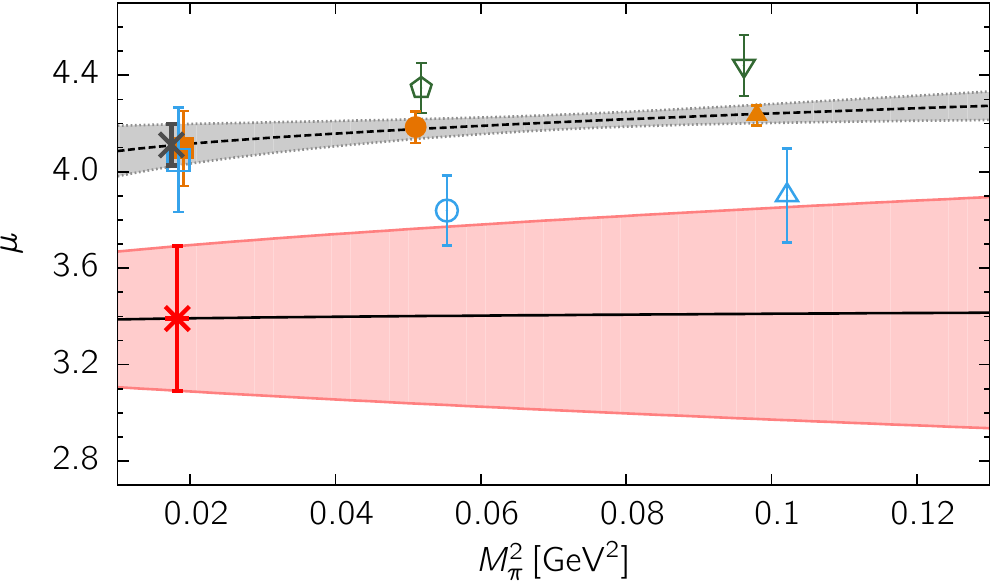}
  %}
    \caption{Chiral continuum extrapolation of $\expv{r_A^2}$,
      $\expv{r_E^2}$, $\expv{r_M^2}$, and $\mu$. The eight data points
      are obtained from the $3^\ast$-state fit for all four
      quantities, followed by $z^{2+4}$ fit for $G_A(Q^2)$ and
      $G_E(Q^2)$, and $z^2$ fit for $\expv{r_M^2}$ and $\mu$. The
      black solid line within the red error band shows the
      extrapolation using Eq.~\eqref{eq:rAsq-extrap} for
      $\expv{r_A^2}$, Eq.~\eqref{eq:rEsq-extrap} for $\expv{r_E^2}$,
      Eq.~\eqref{eq:rMsq-extrap} for $\expv{r_M^2}$ and
      Eq.~\eqref{eq:mu-extrap} for $\mu$. These 4-parameter fits
      (shown versus only $M_\pi^2$) are compared with a fit versus
      only $M_\pi^2$ (setting $c_{2,4}^X=0$ where $(X=A,E,M,\mu)$)
      shown by the black dashed line within the gray error band. The
      red and black crosses at $M_\pi = 135$~MeV are the final
      estimates from these 4 versus 2 parameter fits. }
  \label{fig:ff-extrap}
\end{figure}

In Fig.~\ref{fig:ff-extrap}, we plot the data and compare the chiral
continuum extrapolation of $\expv{r_M^2}$ and $\mu$ for the $z^2$ fit
to the data from the $3^\ast$-state analysis for two cases. The pink
band shows the 4-parameter fit using Eqs.~\eqref{eq:rMsq-extrap}
and~\eqref{eq:mu-extrap} projected on to the $M_\pi^2$ axis, i.e., fit
to the data extrapolated to their continuum values in the other
two variables, $a$ and $M_\pi L$. The grey band shows the fit only versus
$M_\pi^2$, i.e., neglecting lattice spacing and
volume dependence by setting $c_2^M=c_4^M=0$.  The plots show that for a given pion mass, both
$\expv{r_M^2}$ and $\mu$ decrease as the lattice spacing
decreases. The fit keeping all four terms in
Eqs.~\eqref{eq:rMsq-extrap} and~\eqref{eq:mu-extrap} is sensitive to
this trend and thus gives smaller estimates. Ignoring the $a$
dependence, the fit versus just $M_\pi^2$ is controlled by the three
0.09\,$\fm$ ensemble points as they have the smallest errors. It gives
$\expv{r_M^2}=0.74(7)$, which fortuitously agrees with the
experimental value $\expv{r_M^2}=0.74(5)$. However, the corresponding estimate of $\mu=4.11(9)$ is still lower than the experimental value $\mu=4.7058$.
%{On the other hand, our 
%estimate of $\mu=1+\kappa_p-\kappa_n$ is $3/4$ of the 
%very precisely determined experimental values $\mu=4.7058$, and 
%the anomalous magnetic moments of proton $\kappa_p=1.7928$ and of the neutron
%$\kappa_n=-1.9130$~\cite{Patrignani:2016xqp}.} 

Overall, it is the 0.06\,$\fm$ data that controls the large negative
slope in the lattice spacing dependence and leads to an underestimate
of both $\expv{r_M^2}$ and $\mu$. Since the statistical errors in the
data from these three 0.06\,$\fm$ ensembles are the largest, reducing 
them will be the focus of future work.

%Practically, we can not take these $t_\text{skip}$, because
%covariance matrix is ill-determined with these values. The symptom
%could be cured with an increased statistics.

%% For the extrapolation of magnetic moment $\mu$ with
%% Eq.~\eqref{eq:mu-extrap}, including all terms, a weighted average of
%% the two $z$-expansion fits without sumrules gives
%% %
%% % a + mpi + fv3
%% % take positive 100% correlation for average
%% \begin{align}
%%   \mu &= 3.45(23) \quad : z-\text{expansion} + 2\text{-state}\,,\CL
%%   \mu &= 3.47(36) \quad : z-\text{expansion} + 3^\ast\text{-state}\,.
%%   \label{eq:mu}
%% \end{align}
%% % a + mpisq + fv1
%% % take positive 100% correlation for average
%% %\begin{align}
%% %  \mu &= 3.48(20) \quad : z-\text{expansion} + 2\text{-state}\,,\\
%% %  \mu &= 3.50(32) \quad : z-\text{expansion} + 3^\ast\text{-state}\,.
%% %  \label{eq:mu}
%% %\end{align}
%% %
%% 
%% We show two extrapolations
%% using Eq.~\eqref{eq:mu-extrap} and all eight ensemble data from $z^2$
%% fit with $3^\ast$-state analysis in Fig.~\ref{fig:ff-extrap}. An
%% extrapolation with $c_2^\mu=c_4^\mu=0$, which turns off the lattice
%% spacing and volume effect terms, pass right through the three
%% $0.09\,\fm$ points and results a enhanced value $\mu=4.11(9)$ but
%% still shows a discrepancy with the experiment. Again, we can see a
%% downward pattern as lattice spacing vanish, $a\to 0$; exceptions are
%% the two physical ensembles which are on top of each other with large
%% errors.
%% 

\begin{table}[!tbh]
  \centering
  \caption{Final results for the isovector charge radii $r_A, r_E$ and
    $r_M$ in unit of $\fm$ and the corresponding masses
    $\mathcal{M}_A, \mathcal{M}_E$ and $\mathcal{M}_M$ in unit of
    $\GeV$. The magnetic moment $\mu_{p-n} \equiv 1 + \kappa_p
    -\kappa_n$. The results are presented separately for the 2- and
    $3^\ast$-state fits used to control the excited-state contamination
    and the dipole and the $z-$expansion fits to capture the $Q^2$
    behavior. }
  \label{tab:final}
  \renewcommand{\arraystretch}{1.1}
  \begin{tabular}{l|c|cc|cc|cc|c}
  \hline\hline
  \multicolumn{1}{c|}{$Q^2$}     & 3-pt.              & $r_A$  & $\mathcal{M}_A$ & $r_E$  & $\mathcal{M}_E$ & $r_M$  & $\mathcal{M}_M$ & $\mu_{p-n}$  \\
\hline
\multirow{2}{*}{$z$-exp.} & $2^{\phantom{\ast}}$        & 0.44(5)   & 1.56(18)            & 0.70(7)   & 0.98(10)             & 0.86(07)   & 0.80(06)             &  3.45(23)    \\
         & $3^\ast$ & 0.50(6)   & 1.36(17)            & 0.83(9)   & 0.82(08)             & 0.82(10)   & 0.83(10)            &  3.47(36)    \\
\hline
\multirow{2}{*}{dipole}   & $2^{\phantom{\ast}}$        & 0.46(2)   & 1.50(07)             & 0.69(2)   & 0.99(03)             & 0.72(03)   & 0.95(04)             &  3.52(15)    \\
         & $3^\ast$ & 0.51(2)   & 1.34(06)             & 0.79(3)   & 0.87(03)             & 0.77(04)   & 0.89(05)             &  3.72(23)    \\
  \hline\hline
  \end{tabular}
\end{table}

\section{Summary}

We have improved the control over excited-state contamination in the
form factor analysis by including the second excite state in the fits.
The results for $r_A$ and $r_E$ from the $3^\ast$-state fits are
closer to the phenomenological value for both the $z$-expansion and the 
dipole analysis. The $3^\ast$-state fits are about $1\sigma$
($3\sigma$) larger for the $z$-expansion (dipole) fit compared to
the corresponding 2-state fit analysis.

The error from the dipole fits is typically a factor of 2--3 smaller than 
that from the $z$-expansion fits as shown in Table~\ref{tab:final}. Given the 
change in the value between 2- and $3^\ast$-state fits, we consider the error 
estimates using the $z$-expansion more realistic. 

The $z$-expansion with $3^\ast$-state fits give an $r_A=0.50(6)\,\fm$ that is
smaller than the phenomenological estimate
$r_A=0.68(16)\,\fm$~\cite{Meyer:2016oeg}. The results for $r_E=0.83(9)\,\fm$ and
$r_M=0.82(10)\,\fm$ are consistent with phenomenological values $r_E=0.8409(4)\,\fm$ and $r_M=0.86(3)\,\fm$. The outlier is our
estimate of the magnetic moment $\mu$ which is about $3/4$ of the
precisely known experimental value $\mu=4.7058$.

Our plan for the future is to increase the statistics on the two
physical pion mass ensembles and understand why the data for the
three form factors $G_A$, $\tilde{G}_P$ and $G_P$ do not satisfy the
axial Ward identity.

\vskip20pt

\begin{acknowledgement}
  {\textbf{Acknowledgement}} We thank the MILC Collaboration for
  providing the 2+1+1-flavor HISQ lattices. We thank Emanuele Mereghetti for 
  discussions. Simulations were carried
  out on computer facilities of (i) the USQCD Collaboration, which are
  funded by the Office of Science of the U.S. Department of Energy,
  (ii) the National Energy Research Scientific Computing Center, a DOE
  Office of Science User Facility supported by the Office of Science
  of the U.S. Department of Energy under Contract
  No. DE-AC02-05CH11231; (iii) Oak Ridge Leadership Computing Facility
  at the Oak Ridge National Laboratory, which is supported by the Office
  of Science of the U.S.  Department of Energy under Contract
  No. DE-AC05- 00OR22725; (iv) Institutional Computing at Los Alamos
  National Laboratory; and (v) the High Performance Computing Center
  at Michigan State University.  The calculations used the Chroma
  software suite~\cite{Edwards:2004sx}. This work is supported by the
  U.S. Department of Energy, Office of Science of High Energy Physics
  under contract number~DE-KA-1401020 and the LANL LDRD program. The
  work of H-W. Lin was supported in part by the M. Hildred Blewett
  Fellowship of the American Physical Society.
\end{acknowledgement}

\bibliography{lattice2017}

\begin{thebibliography}{12}

\bibitem{Rajan:2017lxk}
R.~Gupta, Y.C. Jang, H.W. Lin, B.~Yoon, B.~Bhattacharya (2017),
  \texttt{1705.06834}

\bibitem{AguilarArevalo:2010zc}
A.A. Aguilar-Arevalo et~al. (MiniBooNE), Phys. Rev. \textbf{D81}, 092005
  (2010), \texttt{1002.2680}

\bibitem{Bernard:1998gv}
V.~Bernard, H.W. Fearing, T.R. Hemmert, U.G. Meissner, Nucl. Phys.
  \textbf{A635}, 121 (1998), [Erratum: Nucl. Phys.A642,563(1998)],
  \texttt{hep-ph/9801297}

\bibitem{Meyer:2016oeg}
A.S. Meyer, M.~Betancourt, R.~Gran, R.J. Hill, Phys. Rev. \textbf{D93}, 113015
  (2016), \texttt{1603.03048}

\bibitem{Bernard:2001rs}
V.~Bernard, L.~Elouadrhiri, U.G. Meissner, J. Phys. \textbf{G28}, R1 (2002),
  \texttt{hep-ph/0107088}

\bibitem{Gockeler:2003ay}
M.~Gockeler, T.R. Hemmert, R.~Horsley, D.~Pleiter, P.E.L. Rakow, A.~Schafer,
  G.~Schierholz (QCDSF), Phys. Rev. \textbf{D71}, 034508 (2005),
  \texttt{hep-lat/0303019}

\bibitem{Mohr:2015ccw}
P.J. Mohr, D.B. Newell, B.N. Taylor, Rev. Mod. Phys. \textbf{88}, 035009
  (2016), \texttt{1507.07956}

\bibitem{Antognini:2015moa}
A.~Antognini et~al., EPJ Web Conf. \textbf{113}, 01006 (2016),
  \texttt{1509.03235}

\bibitem{Beane:2004tw}
S.R. Beane, Phys. Rev. \textbf{D70}, 034507 (2004), \texttt{hep-lat/0403015}

\bibitem{Meissner:1997hn}
U.G. Meissner, S.~Steininger, Nucl. Phys. \textbf{B499}, 349 (1997),
  \texttt{hep-ph/9701260}

\bibitem{Patrignani:2016xqp}
C.~Patrignani et~al. (Particle Data Group), Chin. Phys. \textbf{C40}, 100001
  (2016)

\bibitem{Edwards:2004sx}
R.G. Edwards, B.~Joo (SciDAC Collaboration, LHPC Collaboration, UKQCD
  Collaboration), Nucl.Phys.Proc.Suppl. \textbf{140}, 832 (2005),
  \texttt{hep-lat/0409003}

\end{thebibliography}

\end{document}